\documentclass[12pt,a4paper]{article}
\usepackage{amsfonts,latexsym}
\usepackage{amsmath,amssymb}
\usepackage{graphicx,color}

\renewcommand{\thefootnote}{\fnsymbol{footnote}}

\begin{document}

\vspace{12mm}

\begin{center}
{{{\Large {\bf Instability of Schwarzschild-AdS black hole \\ in  Einstein-Weyl gravity }}}}\\[10mm]

{Yun Soo Myung\footnote{e-mail address: ysmyung@inje.ac.kr}}\\[8mm]

{Institute of Basic Sciences and Department  of Computer Simulation, Inje University Gimhae 621-749, Korea\\[0pt]}

\end{center}
\vspace{2mm}

\begin{abstract}
We  investigate  the classical stability of Schwarzschild-AdS black
hole in a massive gravity theory of the
 Einstein-Weyl  gravity. It turns out that the
linearized Einstein tensor perturbations exhibit unstable modes
featuring the Gregory-Laflamme instability of five-dimensional AdS
black string, in contrast to the stable Schwarzschild-AdS black hole
in  Einstein gravity.  We point out  that the instability of the
black hole in the Einstein-Weyl gravity  arises from  the
massiveness but not a feature of fourth-order derivative theory
giving ghost states.
\end{abstract}
\vspace{5mm}



\vspace{1.5cm}

\hspace{11.5cm}{Typeset Using \LaTeX}
\newpage
\renewcommand{\thefootnote}{\arabic{footnote}}
\setcounter{footnote}{0}


\section{Introduction}

Recently, Babichev and Fabbri~\cite{Babichev:2013una} have shown
that the massive linearized equation around the Schwarzschild black
hole in both de Rham, Gabadadze, and Tolley (dRGT)
theory~\cite{deRham:2010ik} and its bigravity
extension~\cite{Hassan:2011zd} gives rise to an instability of
$s(l=0)$-mode (spherically symmetric mode) with $l$ the spheroidal
harmonic index. This was done by comparing it with the
four-dimensional linearized equation around the five-dimensional
black string where
 the Gregory-Laflamme (GL) instability was found~\cite{Gregory:1993vy}. It turned out
that the bimetric black hole is unstable provided a mass of
$m'=m(1+1/\kappa)^{1/2}$ satisfies a bound of $0<m'<{\cal O}(1)/r_0$
with $r_0$ the horizon radius in the metric function $f(r)=1-r_0/r$.
We note that the limit of $\kappa \to \infty$ recovers the black
hole in the dRGT theory. The black hole in the dRGT theory is also
unstable because $m$ satisfies a bound of $0<m<{\cal O}(1)/r_0$.  In
addition, the authors~\cite{Brito:2013wya} have confirmed this
result by considering the Schwarzschild-de Sitter black hole and
extending the $l=0$ mode to generic modes of $l\not=0$.  These
results may indicate an important fact that the static black holes
do not exist in massive gravity theory.

On the other hand, Whitt~\cite{Whitt:1985ki} has insisted thirty
years ago  that provided both massive spin-0 and spin-2 gravitons
are non-tachyonic, the Schwarzschild black hole is classically
stable in a massive theory of fourth-order gravity  when he uses the
linearized-Ricci tensor equation. In this case, one does not worry
about the ghost instability arising from the fourth-order gravity
theory because the linearized-Ricci tensor satisfies  a second-order
tensor equation. Recently, the author  has   revisited this
stability issue.  As expected, it was  shown that
 the black hole in fourth-order gravity with $\alpha=-3\beta$ is unstable
provided the graviton mass of  $m_2=1/\sqrt{3\beta}$  satisfies  a
bound of $0<m_2<{\cal O}(1)/r_0$~\cite{Myung:2013doa}. This was
performed by comparing the linearized-Ricci tensor equation  with
the four-dimensional metric perturbation equation around the
five-dimensional black string~\cite{Gregory:1993vy}.

In this work, we wish to reexamine  the classical stability of
Schwarzschild-AdS (SAdS) black hole in
 Einstein-Weyl  gravity which was known to be stable against the metric perturbation~\cite{Liu:2011kf}.
By contrast,  it is shown  that solving both the linearized-Einstein
tensor equation and the metric perturbation equation exhibit
unstable modes featuring the GL instability of five-dimensional AdS
black string~\cite{Hirayama:2001bi}.    It confirms that  the GL
instability of the black hole in the Einstein-Weyl gravity is due to
the massiveness but not a feature of fourth-order gravity giving
ghost states.

Taking into account the number of degrees of freedom (DOF), it is
helpful to show why  the SAdS black hole is physically stable in
the Einstein gravity~\cite{Cardoso:2001bb,Ishibashi:2003ap},
whereas the SAdS black hole is unstable in the Einstein-Weyl
gravity. The number of DOF of the metric perturbation is 2 in the
Einstein gravity, while the number of DOF  is 5 in the
Einstein-Weyl gravity. The $s$-mode analysis of the massive
graviton with $5$ DOF shows the GL instability. The $s$-mode
analysis is relevant to the massive graviton in the Einstein-Weyl
gravity but not to the massless graviton in the Einstein gravity.

\section{Linearized Einstein-Weyl gravity}
We start with the fourth-order gravity in AdS$_4$  spacetimes
\begin{eqnarray}
S_{\rm FO}=\frac{1}{16 \pi}\int d^4 x\sqrt{-g} \Big[R-2\Lambda
+\alpha R_{\mu\nu}R^{\mu\nu}+\beta R^2\Big] \label{Action}
\end{eqnarray} with two arbitrary parameters $\alpha$ and $\beta$.
Although this  theory is renormalizable in Minkowski
spacetimes~\cite{stelle},  the massive spin-2 graviton suffers from
having ghosts. A massive spin-0 graviton is decoupled  for the
choice of  $\alpha=-3\beta$, which leads to a critical gravity with
$\beta=-1/2\Lambda$~\cite{LP,Lu:2011ks}. For a higher dimensional
critical gravity, see a reference of \cite{Deser:2011xc}.

 The
Einstein-Weyl gravity is defined under the condition of
$\alpha=-3\beta$ as
\begin{eqnarray}S_{\rm EW}=\frac{1}{16 \pi}\int d^4 x\sqrt{-g}
\Big[R-2\Lambda
-\frac{\beta}{6}C^{\mu\nu\rho\sigma}C_{\mu\nu\rho\sigma}\Big]
\label{EWAction}
\end{eqnarray}
with
\begin{equation}
C^{\mu\nu\rho\sigma}C_{\mu\nu\rho\sigma}=2\Big(R^{\mu\nu}R_{\mu\nu}-\frac{1}{3}R^2\Big)+
(R^{\mu\nu\rho\sigma}R_{\mu\nu\rho\sigma}-4R^{\mu\nu}R_{\mu\nu}+R^2).
\end{equation}
Here the last of Gauss-Bonnet term could be neglected  because it
does not contribute to equation of motion.

From (\ref{Action}),  the Einstein equation is derived to be
\begin{equation} \label{equa1}
G_{\mu\nu}+E_{\mu\nu}=0,
\end{equation}
where the Einstein tensor  is given by \begin{equation}
G_{\mu\nu}=R_{\mu\nu}-\frac{1}{2} Rg_{\mu\nu}+\Lambda g_{\mu\nu}
\end{equation}
and $E_{\mu\nu}$  takes the form
\begin{eqnarray} \label{equa2}
E_{\mu\nu}&=& 2\alpha
\Big(R_{\mu\rho\nu\sigma}R^{\rho\sigma}-\frac{1}{4}
R^{\rho\sigma}R_{\rho\sigma}g_{\mu\nu}\Big)+2\beta
R\Big(R_{\mu\nu}-\frac{1}{4} Rg_{\mu\nu}\Big) \nonumber \\
&+&
\alpha\Big(\nabla^2R_{\mu\nu}+\frac{1}{2}\nabla^2Rg_{\mu\nu}-\nabla_\mu\nabla_\nu
R\Big) +2\beta\Big(g_{\mu\nu} \nabla^2R-\nabla_\mu \nabla_\nu
R\Big).
\end{eqnarray}
It is wellknown that Eq.(\ref{equa1}) provides  the SAdS black
hole solution~\cite{LP,Liu:2011kf} \begin{equation} \label{sch}
ds^2_{\rm SAdS}=\bar{g}_{\mu\nu}dx^\mu
dx^\nu=-f(r)dt^2+\frac{dr^2}{f(r)}+r^2d\Omega^2_2
\end{equation}
with the metric function \begin{equation} \label{num}
f(r)=1-\frac{r_0}{r}-\frac{\Lambda}{3}r^2,~~\Lambda=-\frac{3}{\ell^2}.
\end{equation}
 Here $\ell$ denotes the
curvature radius of AdS$_4$ spacetimes. We note that a mass
parameter of $r_0=r_+(1+r_+^2/\ell^2)$ is not the horizon radius
$r_+$ which is obtained as a solution to $f(r_+)=0$. Hereafter we
denote the background quantities with the ``overbar''. In this
case, the background Ricci tensor is given by
\begin{equation} \label{beeq}
\bar{R}_{\mu\nu}=\Lambda\bar{g}_{\mu\nu}.
\end{equation}
It is easy to show that the SAdS black hole  solution (\ref{sch})
to the Einstein equation of $G_{\mu\nu}=0$  is also the solution
to the Einstein-Weyl gravity when one substitutes (\ref{beeq})
together with $\bar{R}=4\Lambda$ into (\ref{equa2}). To perform
the stability analysis, we usually introduce the metric
perturbation around the SAdS black hole
\begin{eqnarray} \label{m-p}
g_{\mu\nu}=\bar{g}_{\mu\nu}+h_{\mu\nu}.
\end{eqnarray}
Then, the linearized Einstein equation takes the form
\begin{eqnarray} \label{lin-eq}
\Big[1+2\Lambda(\alpha+4\beta)\Big]\delta G
_{\mu\nu}&+&\alpha\Big[\bar{\nabla}^2\delta
G_{\mu\nu}+2\bar{R}_{\rho\mu\sigma\nu}\delta
G^{\rho\sigma}-\frac{2\Lambda}{3} \delta R \bar{g}_{\mu\nu}\Big]
\nonumber \\
&+&(\alpha+2\beta)\Big[-\bar{\nabla}_\mu\bar{\nabla}_\nu+\bar{g}_{\mu\nu}\bar{\nabla}^2
+\Lambda \bar{g}_{\mu\nu}\Big] \delta R=0,
\end{eqnarray}
where the linearized Einstein tensor, Ricci tensor, and Ricci scalar
are given by
\begin{eqnarray}
\delta G_{\mu\nu}&=&\delta R_{\mu\nu}-\frac{1}{2} \delta
R\bar{g}_{\mu\nu}-\Lambda h_{\mu\nu},
\label{ein-t} \\
\delta
R_{\mu\nu}&=&\frac{1}{2}\Big(\bar{\nabla}^{\rho}\bar{\nabla}_{\mu}h_{\nu\rho}+
\bar{\nabla}^{\rho}\bar{\nabla}_{\nu}h_{\mu\rho}-\bar{\nabla}^2h_{\mu\nu}-\bar{\nabla}_{\mu}
\bar{\nabla}_{\nu}h\Big), \label{ricc-t} \\
\delta R&=& \bar{g}^{\mu\nu}\delta
R_{\mu\nu}-h^{\mu\nu}\bar{R}_{\mu\nu}= \bar{\nabla}^\mu
\bar{\nabla}^\nu h_{\mu\nu}-\bar{\nabla}^2 h-\Lambda h
\label{Ricc-s}.
\end{eqnarray}
with $h=h^\rho~_\rho$. It is very difficult to solve the linearized
equation (\ref{lin-eq}) directly because it is a coupled
second-order equation for $\delta G_{\mu\nu}$ and $\delta R$. Thus,
we attempt  to decouple $\delta R$ from (\ref{lin-eq}).

For this purpose, we take  the trace of (\ref{lin-eq}) which leads
to
\begin{equation}
\Big[4(\alpha+3\beta)\bar{\nabla}^2-2\Big]\delta R=0.
\end{equation}
It implies that the D'Alembertian operator could be  removed if one
chooses \begin{equation} \alpha=-3\beta. \end{equation}
In this case,
the linearized Ricci scalar is constrained to vanish
\begin{equation}
\delta R=0.
\end{equation}
Plugging $\delta R=0$  into Eq. (\ref{lin-eq}) leads to the equation
for the linearized Einstein tensor solely
\begin{equation} \label{slin-eq}
\Big(\bar{\nabla}^2-\frac{2\Lambda}{3}-\frac{1}{3\beta}\Big)\delta
G_{\mu\nu}+ 2\bar{R}_{\rho\mu\sigma\nu}\delta G^{\rho\sigma}=0.
\end{equation}
This shows clearly  why we consider the Einstein-Weyl gravity
(\ref{EWAction}) with $\alpha=-3\beta$ instead of the fourth-order
gravity action (\ref{Action}) with arbitrary $\alpha$ and $\beta$.

Before we proceed, we wish to mention that the metric perturbation
is not suitable  for analyzing the SAdS black hole stability in the
Einstein-Weyl gravity.  For simplicity, we consider the AdS$_4$
spacetimes background whose curvature tensor takes a simple  form
\begin{equation}
\bar{R}_{\mu\nu\rho\sigma}=\frac{\Lambda}{3}(\bar{g}_{\mu\rho}\bar{g}_{\nu\sigma}-\bar{g}_{\mu\sigma}\bar{g}_{\nu\rho}).
\end{equation}
 After choosing the transverse-traceless
gauge (TTG) \begin{equation}\label{ttg}\bar{\nabla}^\mu
h_{\mu\nu}=0~{\rm and}~h=0,
\end{equation}
 Eq. (\ref{slin-eq}) leads
to a fourth-order differential equation~\cite{LP}
\begin{equation} \label{four-eq}
\Big(\bar{\nabla}^2-\frac{2\Lambda}{3}\Big)\Big(\bar{\nabla}^2-\frac{4\Lambda}{3}-\frac{1}{3\beta}\Big)h_{\mu\nu}=0
\end{equation}
which may imply a massless spin-2 graviton equation
\begin{equation} \label{se1-eq}
\Big(\bar{\nabla}^2-\frac{2\Lambda}{3}\Big)h^{m}_{\mu\nu}=0
\end{equation}
 and a massive spin-2 graviton equation
 \begin{equation} \label{se2-eq}
\Big(\bar{\nabla}^2-\frac{2\Lambda}{3}-M^2\Big)h^{M}_{\mu\nu}=0.
\end{equation}
Here the mass squared is given by
\begin{equation}
M^2=\frac{2\Lambda}{3}+\frac{1}{3\beta}=\frac{1}{3\beta}(1+2\beta\Lambda).
\end{equation}
In AdS$_4$ spacetimes, the stability condition  is given by the
absence of tachyonic instability ($M^2 \ge 0$)~\cite{Liu:2011kf},
which implies that $\beta$ must satisfy
\begin{equation}
0<\beta\le -\frac{1}{2\Lambda}=\frac{\ell^2}{6}.
\end{equation}
 In the
massless case of $\beta=-1/2\Lambda=\frac{\ell^2}{6}$,
Eq.(\ref{four-eq}) leads to that for a critical gravity
\begin{equation} \label{crit-eq}
\Big(\bar{\nabla}^2-\frac{2\Lambda}{3}\Big)^4h^{\rm
log}_{\mu\nu}=0,~~\Big(\bar{\nabla}^2-\frac{2\Lambda}{3}\Big)^2h^{\rm
log}_{\mu\nu}=-h^{ m}_{\mu\nu}.
\end{equation}
However,  it was shown that a general mode of $h_{\mu\nu}=c_1h^{\rm
log}_{\mu\nu}+c_2h^{ m}_{\mu\nu}$ suffers from negative norm states
unless one truncates out the log-mode by imposing appropriate
AdS$_4$ boundary conditions.  Up to now, there is no consistent
truncation mechanism to eliminate the log-mode. We recall that this
problem arises because we work with the fourth-order derivative
equation (\ref{four-eq}) for the metric perturbation. In this work,
we do not consider a new unitary gravity for $-\frac{9}{4\ell^2} \le
M^2 <0$~\cite{Lu:2011ks,Hyun:2011ej} because it has still a
non-unitarity problem like the critical gravity.

Going back to the SAdS black hole (\ref{sch}), we rewrite Eq.
(\ref{slin-eq}) as a second-order equation for the linearized
Einstein tensor
\begin{equation} \label{se1m-eq}
\bar{\nabla}^2\delta G_{\mu\nu}+2\bar{R}_{\rho\mu\sigma\nu}\delta
G^{\rho\sigma}=M^2 \delta G_{\mu\nu}.
\end{equation}
If one introduces the Lichnerowicz operator
\begin{equation}
\Delta_{\rm L} \delta G_{\mu\nu}=-\bar{\nabla}^2\delta
G_{\mu\nu}-2\bar{R}_{\rho\mu\sigma\nu}\delta
G^{\rho\sigma}+2\Lambda\delta G_{\mu\nu},\end{equation} the
corresponding equation could be rewritten as
\begin{equation} \label{g-eq}
(\Delta_{\rm L} -2\Lambda+M^2)\delta G_{\mu\nu}=0.
\end{equation}
Taking into account the TTG (\ref{ttg}), the linearized Einstein
tensor reduces to
\begin{equation}
\delta G_{\mu\nu}=-\frac{1}{2}(\Delta_{\rm L}-2\Lambda)h_{\mu\nu}.
\end{equation}
Then, one can rewrite (\ref{g-eq}) as a fourth-order equation for
$h_{\mu\nu}$ \begin{equation} \label{four-eqq}(\Delta_{\rm
L}-2\Lambda+M^2)(\Delta_{\rm L}-2\Lambda)h_{\mu\nu}=0
\end{equation}
which is similar to (\ref{four-eq}) in AdS$_4$ spacetimes. Eq.
(\ref{four-eqq}) may imply a linearized massless equation around the
SAdS black hole~\cite{Liu:2011kf}
 \begin{equation} \label{lem1-eq}
\bar{\nabla}^2h^m_{\mu\nu}+2\bar{R}_{\rho\mu\sigma\nu}h^{m\rho\sigma}=0.
\end{equation}
and  a linearized massive equation for $h^M_{\mu\nu}$
\begin{equation} \label{lem2-eq}
\bar{\nabla}^2h^M_{\mu\nu}+2\bar{R}_{\rho\mu\sigma\nu}h^{M\rho\sigma}=M^2
h^M_{\mu\nu}.
\end{equation}

 At this stage, we wish to point
out the difference between (\ref{se1m-eq}) and (\ref{lem2-eq}). The
former equation is a second-order  equation for the linearized
Einstein tensor, whereas the latter is a suggesting second-order
equation from the fourth-order equation (\ref{four-eqq}) for the
metric perturbation. It is  known that the introduction of
fourth-order derivative terms gives rise to ghost-like massive
graviton~\cite{stelle}, which  may imply an instability of a black
hole even if a black hole solution exists.    Hence, even though
(\ref{se1-eq})[(\ref{se2-eq})] were frequently used as a linearized
massless [massive] equation around the  AdS$_4$
spacetimes~\cite{Liu:2011kf,LP,Lu:2011ks,Deser:2011xc,Hyun:2011ej},
their  validity is not yet  proved because they are free from ghost
states. In order to check  whether (\ref{se2-eq})[(\ref{lem2-eq})]
are reliable  or not, we note that our action (\ref{EWAction})
reveals ghosts when we perform the metric perturbation $h_{\mu\nu}$
around the Minkowski spacetimes with $\Lambda=0$~\cite{stelle}.  Eq.
(\ref{four-eq})[(\ref{four-eqq})] take the form in the Minkowski
background~\cite{Myung:2011nn}
 \begin{equation} \label{source}
 \square\Big(\square-m^2_2\Big)h_{\mu\nu} =-T_{\mu\nu},~~m_2^2=\frac{1}{3\beta}
\end{equation}
with an external source $T_{\mu\nu}$. Replacing $\square$ by $-p^2$,
the metric perturbation is given by
\begin{equation}\label{h-s}
h_{\mu\nu} \sim \frac{T_{\mu\nu}}{p^2}-\frac{T_{\mu\nu}}{p^2+m^2_2}
\end{equation}
which the last  term  spoils the unitarity.  Hence, splitting
(\ref{four-eq})[(\ref{four-eqq})] into two second-order equations
(\ref{se1-eq})[(\ref{lem1-eq})] and (\ref{se2-eq})[(\ref{lem2-eq})]
is dangerous because the `$-$' sign in the front of
(\ref{se2-eq})[(\ref{lem2-eq})] is missed.   As is shown in
(\ref{h-s}), the ghost  arises from this sign when one  performs the
partial fraction.   To this end, the authors in~\cite{Liu:2011kf}
have found the two on-shell energies on the AdS$_4$ spacetime
background
\begin{eqnarray}
E_{m}&=&-\frac{3\beta M^2}{2T}\int
d^4x\sqrt{-\bar{g}}(\nabla^0h^{m\mu\nu})\dot{h}^{m}_{\mu\nu}>0, \\
E_{M}&=&\frac{3\beta M^2}{2T}\int
d^4x\sqrt{-\bar{g}}(\nabla^0h^{M\mu\nu})\dot{h}^{M}_{\mu\nu}<0
\end{eqnarray}
when they compute  each  Hamiltonian which satisfies (\ref{se1-eq})
and (\ref{se2-eq}), respectively.  Thus, for $M^2\not=0$, ghost-like
massive excitation is not avoidable. In order for the theory to be
free from ghosts, one needs to choose $M^2=0(\beta=-1/2\Lambda$)
which corresponds to the critical gravity  where a massive graviton
becomes a massless graviton.   Because of a  missing of `$-$' sign,
we may insist that Eq.(\ref{se2-eq})[(\ref{lem2-eq})] by itself do
not represent a correct linearized equation for studying the
stability of the SAdS black hole in the Einstein-Weyl gravity.
However, the overall `$-$' sign in (\ref{se2-eq})[(\ref{lem2-eq})]
does not make any difference unless an external source is introduced
in the right-hand side as Eq.(\ref{source}) does indicate.
Therefore, the fourth-order gravity does not automatically imply the
instability of the black hole even if one uses (\ref{lem2-eq}).
Hopefully, if one uses (\ref{se1m-eq}) instead of (\ref{lem2-eq}),
one is free from the ghost issue   because (\ref{se1m-eq}) is a
genuine second-order equation.

\section{SAdS black hole
stability in Einstein-Weyl gravity}

In  Einstein gravity, the linearized equation around the
Schwarzschild black hole is given by $\delta R_{\mu\nu}(h)=0$ with
$\delta R_{\mu\nu}(h)$ (\ref{ricc-t}). Then, the metric perturbation
$h_{\mu\nu}$ is classified depending on the transformation
properties under parity, namely odd  and even. Using the
Regge-Wheeler~\cite{Regge:1957td} and Zerilli gauge~\cite{Zeri}, one
obtains two distinct perturbations: odd with 2 DOF and even with 4
DOF. This implies that one starts with 6 DOF after choosing the
Regge-Wheleer gauge, leading to 2  DOF (1 for odd and 1 for even)
for a  massless spin-2 graviton propagation. The Schwarzschild black
hole is stable against the metric perturbation~\cite{Vish,chan}.
Performing the stability analysis of the SAdS black hole in Einstein
gravity, one has to use the linearized equation
\begin{equation} \label{einstein-eq}
\delta G_{\mu\nu}(h)= \delta
R_{\mu\nu}(h)-\frac{\bar{g}_{\mu\nu}}{2} \delta R(h)-\Lambda
h_{\mu\nu}=0,
\end{equation}
which was tuned out to be stable by following the Regge-Wheeler
prescription~\cite{Cardoso:2001bb,Ishibashi:2003ap,Moon:2011sz}. In
these cases, the $s(l=0)$-mode analysis is not necessary to show the
stability of the Schwarzschild  and SAdS black holes because the
massless spin-2 graviton requires modes with $l \ge 2$.

 However, the
$s$-mode  analysis is responsible  for detecting  an instability of
a massive graviton propagating on the SAdS black hole in
Einstein-Weyl gravity. The even-parity metric perturbation is
designed for a single $s$-mode analysis in the massive gravity and
whose form is given by $H_{tt},~H_{tr},~H_{rr},$ and $K$ as
\begin{eqnarray}
h^e_{\mu\nu}=e^{\Omega t} \left(
\begin{array}{cccc}
H_{tt}(r) & H_{tr}(r) & 0 & 0 \cr H_{tr}(r) & H_{rr}(r) & 0 & 0 \cr
0 & 0 &  K(r) & 0 \cr 0 & 0 & 0 & \sin^2\theta K(r)
\end{array}
\right). \label{evenp}
\end{eqnarray}
Even though one starts with 4 DOF, they are  related to each other
when one uses the TTG (\ref{ttg}). Hence, we expect to have one
decoupled equation for  $H_{tr}$.

 For a massive gravity theory
in the Minkowski background, there is correspondence between
linearized Ricci tensor $\delta R_{\mu\nu}$  and Ricci spinor
$\Phi_{ABCD}$ when using the Newman-Penrose
formalism~\cite{Newman:1961qr}.  Here the null real tetrad is
necessary to specify polarization modes of a massive graviton, as
the
 massive gravity  requires null complex tetrad to
specify six polarization modes~\cite{Eardley:1974nw,Moon:2011gg}.
This implies that in fourth-order gravity theory, one may take the
linearized Ricci tensor $\delta R_{\mu\nu}$ (\ref{ricc-t}) with 6
DOF  as physical observables~\cite{Myung:2013doa}. Requiring $\delta
R=0$ further, the DOF of $\delta R_{\mu\nu}$ is five which is the
same DOF for the metric perturbation $h_{\mu\nu}$ in massive gravity
theory.

At this stage, we stress again that  (\ref{se1m-eq}) is considered
as the second-order equation with respect to $\delta G_{\mu\nu}$,
but not the fourth-order equation (\ref{four-eq})  for $h_{\mu\nu}$.
Hence, we propose $\delta G_{\mu\nu}$ as physical observables
propagating  on the SAdS black hole background  instead of $\delta
R_{\mu\nu}$ on the Schwarzschild black hole background.  Also, we
have the tracelessness of $ \delta G^{\mu}~_\mu=-\delta R=0$ and the
transversality  of $\bar{\nabla}^\mu \delta G_{\mu\nu}=0$ from the
contracted Bianchi identity. Then, $\delta G_{\mu\nu}$  describe
exactly five DOF propagating  on the SAdS black hole background
without ghosts.

Since Eq.(\ref{lem2-eq}) is the same linearized equation for
four-dimensional metric perturbation around five-dimensional black
string, we follow  the GL instability analysis  in AdS$_4$
spacetimes~\cite{Hirayama:2001bi}. Eliminating all but $H_{tr}$,
Eq.(\ref{lem2-eq}) reduces to a second-order equation for $H_{tr}$
\begin{equation} \label{second-eq} A(r;r_0,\ell,\Omega^2,M^2)
\frac{d^2}{dr^2}H_{tr} +B\frac{d}{dr}H_{tr}+CH_{tr}=0,
\end{equation}
where $A,B$ and $C$ were given by (20)
in~\cite{Hirayama:2001bi,Kang:2002hx}. We stress again that the
$s$-mode perturbation is described by single DOF but not 5 DOF.  The
authors in~\cite{Hirayama:2001bi} have solved (\ref{second-eq})
numerically and found unstable modes for $0<m<\frac{{\cal
O}(1)}{r_0}$.  See Fig. 1 that is  generated   from the numerical
analysis.  We note that $r_+=1,2,3$ correspond to
$r_0=1.01,2.08,4.64$, respectively. From the observation of Fig. 1
with ${\cal O}(1)\simeq 0.85$, we find unstable modes for
\begin{equation} \label{unst-con}
0<M<\frac{{\cal O}(1)}{r_0} \end{equation} with the mass
\begin{equation} M=\sqrt{\frac{1}{3\beta}-\frac{2}{l^2}}.
\end{equation}
\begin{figure*}[t!]
   \centering
   \includegraphics{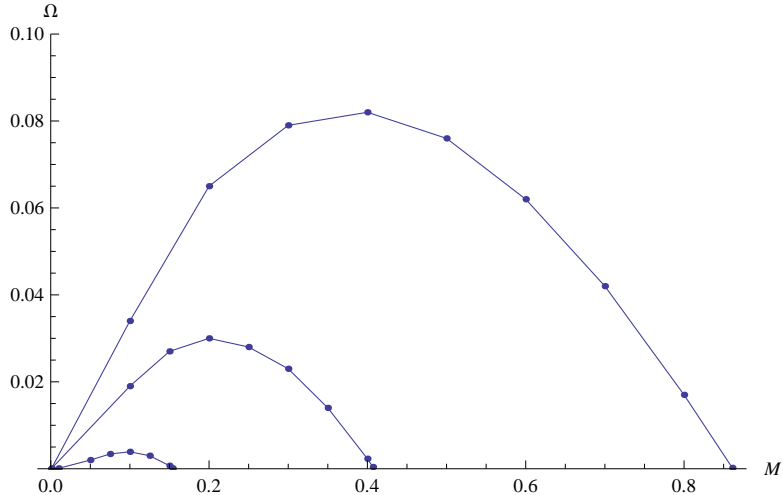}
\caption{Plots of unstable modes on three curves with $r_+=1,2,4$
and $l=10$. The $y(x)$-axis denote $\Omega(M)$. The smallest curve
represents $r_+=4$, the medium denotes $r_+=2$, and the largest one
shows $r_+=1$.  }
\end{figure*}
As the horizon size $r_+$ increases, the instability becomes weak as
in the Schwarzschild black hole~\cite{Kang:2002hx}.

Similarly, we find  Eq.(\ref{lem2-eq}) when we  replace    $ \delta
G_{\mu\nu}$ by  $h_{\mu\nu}$  in (\ref{se1m-eq}).  Hence, a relevant
equation for $\delta G_{tr}$ takes the same form
\begin{equation} \label{secondG-eq} A(r;r_0,\ell,\Omega^2,M^2)
\frac{d^2}{dr^2}\delta G_{tr} +B\frac{d}{dr}\delta G_{tr}+C\delta
G_{tr}=0
\end{equation}
which shows the same unstable modes appeared in Fig. 1.
This implies
that  even if one uses (\ref{lem2-eq}) as a linearized massive
equation~\cite{Liu:2011kf}, our conclusion remains unchanged because
(\ref{lem2-eq}) and (\ref{se1m-eq}) are the same equation for
different tensors.

Consequently,  the instability arises  from the massiveness ($M>0$)
but not from a feature of the fourth-order equation which gives the
`$-$' sign (ghost=negative norm state) when one splits  it into two
second-order equations.  This implies that static black holes in
massive gravity theory do not exist and/or they do not form in the
gravitational collapse. If a black hole was formed in the massive
gravity theory, one may ask what is the end-state of such
instability. For unstable black strings of SAdS$\times$R with
translational symmetry, there are some evidences that break-up
occurs~\cite{Lehner:2010pn}. However,  we consider  a spherically
symmetric black hole in four dimensions.  A possible end-state may
be a spherically symmetric black hole endowed with a graviton
cloud~\cite{Brito:2013wya,Volkov:2012wp}

Finally, in the case of $M=0(\beta=-1/2\Lambda)$, the theory becomes
massless and is stable against the Einstein tensor perturbation.
However, this corresponds precisely to the critical gravity when one
uses  the metric perturbation.  Here we  have a non-unitarity issue
due to the log-mode like (\ref{crit-eq}).  Also,  one finds that
${\cal M}$(ADT mass)=0 and $S$(Wald's entropy)=0 at the critical
point, leading to a vacuum but not a black hole~\cite{LP}.

 \vspace{1cm}

{\bf Acknowledgments}
 \vspace{1cm}

The author thanks Taeyoon Moon for drawing Figure.  This work was
supported by the National Research Foundation of Korea (NRF) grant
funded by the Korea government (MEST) (No.2012-R1A1A2A10040499).


\begin{thebibliography}{99}




\bibitem{Babichev:2013una}
  E.~Babichev and A.~Fabbri,
   Class.\ Quant.\ Grav.\  {\bf 30}, 152001 (2013)  [arXiv:1304.5992 [gr-qc]].

\bibitem{deRham:2010ik}
  C.~de Rham and G.~Gabadadze,
  Phys.\ Rev.\ D {\bf 82}, 044020 (2010)  [arXiv:1007.0443 [hep-th]].

\bibitem{Hassan:2011zd}
  S.~F.~Hassan and R.~A.~Rosen,
  JHEP {\bf 1202}, 126 (2012)  [arXiv:1109.3515 [hep-th]].


\bibitem{Gregory:1993vy}
  R.~Gregory and R.~Laflamme,
   Phys.\ Rev.\ Lett.\  {\bf 70}, 2837 (1993)  [hep-th/9301052].



\bibitem{Brito:2013wya}
  R.~Brito, V.~Cardoso and P.~Pani,
  Phys.\  Rev.\ D {\bf 88}, 023514 (2013)  [arXiv:1304.6725 [gr-qc]].


\bibitem{Whitt:1985ki}
  B.~Whitt,
   Phys.\ Rev.\ D {\bf 32}, 379 (1985).


\bibitem{Myung:2013doa}
  Y.~S.~Myung,
 Phys.\  Rev.\ D {\bf 88}, 024039 (2013)  [arXiv:1306.3725 [gr-qc]].

\bibitem{Liu:2011kf}
  H.~Liu, H.~Lu and M.~Luo,
   Int.\ J.\ Mod.\ Phys.\ D {\bf 21}, 1250020 (2012)  [arXiv:1104.2623 [hep-th]].


\bibitem{Hirayama:2001bi}
  T.~Hirayama and G.~Kang,
  Phys.\ Rev.\ D {\bf 64}, 064010 (2001)  [hep-th/0104213].


\bibitem{Cardoso:2001bb}
  V.~Cardoso and J.~P.~S.~Lemos,
  Phys.\ Rev.\ D {\bf 64}, 084017 (2001)  [gr-qc/0105103].

\bibitem{Ishibashi:2003ap}
  A.~Ishibashi and H.~Kodama,
   Prog.\ Theor.\ Phys.\  {\bf 110}, 901 (2003)  [hep-th/0305185].

\bibitem{stelle}
  K.~S.~Stelle,
  Phys.\ Rev.\  D {\bf 16}, 953 (1977).



\bibitem{LP}
 H.~Lu and C.~N.~Pope,
  Phys.\ Rev.\ Lett.\  {\bf 106}, 181302 (2011)
  [arXiv:1101.1971 [hep-th]].

\bibitem{Lu:2011ks}
  H.~Lu, Y.~Pang and C.~N.~Pope,
  Phys.\ Rev.\ D {\bf 84}, 064001 (2011)  [arXiv:1106.4657 [hep-th]].



\bibitem{Deser:2011xc}
  S.~Deser, H.~Liu, H.~Lu, C.~N.~Pope, T.~C.~Sisman and B.~Tekin,
  Phys.\ Rev.\ D {\bf 83}, 061502 (2011)  [arXiv:1101.4009 [hep-th]].

\bibitem{Hyun:2011ej}
  S.~-J.~Hyun, W.~-J.~Jang, J.~-H.~Jeong and S.~-H.~Yi,
   JHEP {\bf 1201}, 054 (2012)  [arXiv:1111.1175 [hep-th]].


\bibitem{Myung:2011nn}
  Y.~S.~Myung,
   arXiv:1107.3594 [hep-th].






\bibitem{Regge:1957td}
  T.~Regge and J.~A.~Wheeler,
   Phys.\ Rev.\  {\bf 108}, 1063 (1957).

\bibitem{Zeri}
  F.~J.~Zerilli,
  Phys.\ Rev.\ Lett.\  {\bf 24}, 737 (1970).

\bibitem{Vish}
  C.~V.~Vishveshwara,
  Phys.\ Rev.\  D {\bf 1}, 2870 (1970).


\bibitem{chan} S. Chandrasekhar, in The Mathematical Theory of Black Holes
(Oxford University, New York, 1983).


\bibitem{Moon:2011sz}
  T.~Moon, Y.~S.~Myung and E.~J.~Son,
  Eur.\ Phys.\ J.\ C {\bf 71}, 1777 (2011)  [arXiv:1104.1908 [gr-qc]].





\bibitem{Newman:1961qr}
  E.~Newman and R.~Penrose,
  J.\ Math.\ Phys.\  {\bf 3}, 566 (1962).



\bibitem{Eardley:1974nw}
  D.~M.~Eardley, D.~L.~Lee and A.~P.~Lightman,
   Phys.\ Rev.\ D {\bf 8}, 3308 (1973).

\bibitem{Moon:2011gg}
  T.~Moon and Y.~S.~Myung,
   Phys.\ Rev.\ D {\bf 85}, 027501 (2012)  [arXiv:1111.2196 [gr-qc]].


\bibitem{Kang:2002hx}
  G.~Kang,
  hep-th/0202147.

\bibitem{Lehner:2010pn}
  L.~Lehner and F.~Pretorius,
  Phys.\ Rev.\ Lett.\  {\bf 105}, 101102 (2010)  [arXiv:1006.5960 [hep-th]].


\bibitem{Volkov:2012wp}
  M.~S.~Volkov,
  Phys.\ Rev.\ D {\bf 85}, 124043 (2012)  [arXiv:1202.6682 [hep-th]].


\end{thebibliography}
\end{document}